\documentclass[12pt]{article}
\usepackage{amssymb}
\usepackage{amsmath}

\input{tcilatex}
\begin{document}

\title{\textbf{Spatially Homogeneous Rotating Solution in f(R) Gravity and
Its Energy Contents}}
\author{M. Jamil Amir \thanks{%
mjamil.dgk@gmail.com} and Saima Naheed \thanks{%
symanahid@yahoo.com} \\
Department of Mathematics, University of Sargodha,\\
Sargodha-40100, Pakistan.}
\date{}
\maketitle

\begin{abstract}
In this paper, the metric approach of $f(R)$ theory of gravity is
used to investigate the exact vacuum solutions of spatially
homogeneous rotating spacetimes. For this purpose, $R$ is replaced
by $f(R)$ in the standard Einstein-Hilbert action and the set of
modified Einstein field equations reduce to a single equation. We
adopt the assumption of constant Ricci scalar which maybe zero or
non-zero. Moreover, the energy density  of the non-trivial
solution has been evaluated by using the generalized
Landau-Lifshitz energy-momentum complex in the perspective of
$f(R)$ gravity for some appropriate $f(R)$ model, which turns out
to be a constant quantity.
\end{abstract}


\section{Introduction}

In recent years, the attention towards modified theories of
gravity is increasing rapidly. Although General Relativity(GR) has
solved many problems but yet there are some important issues like,
the accelerated expansion of the universe \cite {[1]noijri},
cosmological constants or dark energy problem: \ the $120$ orders
of magnitude variation between the theoretical and observational
values of the vacuum energy density, which have to be explored.
Lovelock and $f(R)$ theories are some efforts to modify the GR, in
which the higher powers of R and its derivatives are used to solve
the unresolved issues.

Eddington \cite{[2]eddington} was the first who pondered the
$f(R)$ actions. Afterwards Buchdahl \cite{[3]buchdahl} studied
these $f(R)$ actions in case of non-singular oscillating
cosmologies. These theories are termed as higher order
gravitational theories and proved that these models are equivalent
to scalar-tensor models of gravity. It is palpable that firstly we
must check their consistency with the solar system tests of
Einstein gravity \cite{[4]navarro}. It was observed that without
bringing new degree of freedom, it is not possible to perform
these tests and meanwhile to account for the accelerated expansion
of the universe, in most of the models. Soon after, Sawicki
\cite{[5]sawicki} and Starobinski \cite{[6]starobinski} have
illustrated that without bringing new degrees of freedom one can
account for both the solar system tests and the accelerated
expansion of the universe at the same time.

The most spaciously discovered exact solutions in $f(R)$ gravity
are the spherically symmetric solutions which were found by
Multam$\ddot{a}$ki and Vilja \cite{[7]ft22}. They proved that the
whole set of field equations in $f(R)$\ gravity yields exactly the
Schwarzschild de Sitter metric. Lukas and Francisco \cite{[8]ft27}
have analyzed the exact solutions of static spherically symmetric
spacetimes in $f(R)$ modified theories of gravity. Cylindrical
symmetric solution in $f(R)$ theory were
 explored by Momeni, Azadi
and Nouri-Zonoz \cite{[9]ft25}. This work was extended by Momeni
and Gholizade \cite{[10]ft26} to the general cylindrical symmetric
solutions.

Sharif and Farasat studied plane symmetric static
solution \cite{[11]farasat paper refrence} and vacuum solutions of
Bianchi types $I$ and $V$ spacetimes \cite{[12]f paper 2} in
$f(R)$ theory of gravity by using metric affine approach. They
also extended this work to perfect fluid solutions and found the
non-vacuum solutions of Bianchi types $I$ and $V$ in $f(R)$
gravity by taking stiff matter \cite{[13]f paper 3}. Farast also studied Bianchi type $I$, $III$ and Kantowski-
Sachs spacetimes in $f(R)$ gravity \cite{[13a]f paper 330} and \cite{[13b]f paper 50}. Black hole solutions in $F(R)$ gravity
with conformal anomaly are found by Hendi and Momeni \cite{[13c]Hendi paper 71}.
Noether symmetry approach in $f(R)$-Tachyon model is discussed in \cite{[13d] paper B702}.
 Recently Farasat et al. studied conserved quantities in $f(R)$ gravity
via Noether symmetry \cite{[13e]f paper 29(8)}.

Since the Einstein era the localization of energy has been a complicated and
controversial issue in GR which is still unanswerable.  Einstein
\cite{[14]11of m.jamil} initiated the energy-momentum
pseudo-tensor and gave the energy-momentum conservation laws as
follows
\begin{eqnarray}
\frac{\partial}{\partial
x^\nu}\{\sqrt{-g}(T_{\mu}^{\nu}+t_{\mu}^{\nu})\}=0,
~~~~~~~~~~~(\mu, \nu =0,1,2,3)\nonumber
\end{eqnarray}
where $t_{\mu }^{\nu}$ is energy-momentum density of gravitation
and $ T_{\mu}^{\nu}$ is energy-momentum density of matter.
Fundamental nature of conservation laws was deeply explained by
Bergmann \cite{[15]farast 1}-\cite{[17]farast 3}.

By considering the geodesic coordinate system, Landau-Lifshitz
\cite{[18]farast 2. ref 6} derived  the energy-momentum complex
(EMC) at some particular point of a spacetime. Various authors,
like, Tolman \cite{[19]farast 2. ref 7}, Papapetrou
\cite{[20]farast 2. ref 8} Bergmann \cite{[21]farast 2. ref 9},
Goldberg \cite{[22]farast 2. ref 10}, M\"{O}ller \cite{[23]farast
2. ref 11} and Weinberg \cite{[24]farast 2. ref 12} added their
own EMCs. Misner et al. \cite{[25]farast 2. ref 13} proved that
energy can only be localized in spherical systems. But soon after,
Cooperstock and Sarracino \cite{[26]farast 2. ref 14} proved that
if energy is localizable for spherical systems, then it can be
localized (to express it as a unique tensor quantity) in any
system.

Multam\"{a}k et al. \cite{[27]farast 2. ref 32} generalized the
Landau-Lifshitz EMC in the context of $f(R)$ theory of gravity. In
the framework of metric $f(R)$ gravity, the energy-momentum
distribution (EMD) may be evaluated by using the generalized
Landau-Lifshitz prescription. They calculated the energy density
for the Schwarzschild de Sitter spacetime. Recently, Sharif and
Farasat \cite{[28]sharif paper} calculated the energy density of
plane symmetric static solutions and cosmic string spacetime by
using generalized Landau-Lifshitz prescription.

Valerio Faraoni and Shahn Nadeau \cite{[29]ref f (r) models paper}
have discussed some important $f(R)$ models along with their
stability conditions. We are now extending this work by exploring
the spatially homogeneous rotating solutions in $f(R)$ gravity and
the energy contents of the obtained non-trivial solution for
particular $f(R)$ model.

The paper is arranged as follows: In section $2$, we have briefly
discussed the modified field equations in metric approach of
$f(R)$ gravity. Section $3$ contains the derivation of generalized
Landau-Lifshitz EMC. The exact vacuum solution, by applying the
constant curvature condition, of spatially homogeneous rotating
spacetimes is described in section $4$. In section $5$, we have
calculated the energy density of the solution obtained in previous
section by using Landau-Lifshitz EMC. We summarized the results in
the last section.

\section{Field Equations in f(R) Theory of Gravity}
In this section, we derive the field equations. For this purpose,
we used the metric approach of $f(R)$ theory of gravity. In this
approach, the variation of the action is done with respect to the
metric tensor only. The action of $f(R)$ theory is stated as
\begin{equation}
S=\int \sqrt{-g}(\frac{1}{16\pi G}f(R)+L_{m})d^{4}x,  \label{1}
\end{equation}%
where $f(R)$ is a general function of Ricci scalar $R$ and $L_{m}$
is the matter Lagrangian. The replacement of $R$ by $f(R)$ in the
standard Einstien-Hilbert action gives us this action. The
corresponding field equations can be obtained by varying this
action with respect to metric tensor $g_{\mu \nu }$ and are given
by
\begin{equation}
F(R)R_{\mu\nu}-\frac{1}{2}f(R)g_{\mu\nu}-\nabla_{\mu}\nabla_{\nu}F(R)+g_{\mu\nu}\Box
F(R)=\kappa T_{\mu\nu}, \label{2}
\end{equation}
where
\begin{equation}
F(R)=\frac{df(R)}{dR},~~~~~~~ \Box =\nabla^{\mu}\nabla_{\mu}.
\label{3}
\end{equation}
Here, $\nabla _{\mu }$ represents the covariant derivative, $\Box
=\nabla ^{\mu }\nabla _{\mu }$ is called D' Alembert operator and
$T_{\mu \nu }$ is the standard matter energy-momentum tensor
derived from $L_{m}$. These are the fourth order partial
differential equations in the metric tensor. For $f(R)=R$ these
equations reduce to the famous Einstein field equations of GR.
After contraction the field equations turn out to be
\begin{equation}
F(R)R-2f(R)+3\Box F(R)=\kappa T   \label{4}
\end{equation}
and, in vacuum, i.e., when $T=0$, the last equation takes the
form
\begin{equation} F(R)R-2f(R)+3\square F(R)=0.  \label{5}
\end{equation}
This gives an important relationship between $f(R)$ and $F(R)$
which will be helpful to simplify the field equations and to determine the $%
f(R)$. It is to be noted that any metric with constant scalar
curvature, say $R=R_{0}$, is a solution of the Eq.(\ref{5}) if the
following equation holds
\begin{equation}
F(R_{0})R_{0}-2f(R_{0})=0.  \label{6}
\end{equation}
This is the constant scalar curvature condition in vacuum and for
non-vacuum case, it takes the form
\begin{equation}
F(R_{0})R_{0}-2f(R_{0})=\kappa T.  \label{7}
\end{equation}
These conditions play an important role to find the acceptability of $f(R)$
models.

\section{Generaized Landau-Lifshitz Energy-\newline
Momentum Complex}

The generalized Landau-Lifshits EMC is given by
\begin{equation}
\tau ^{\mu \nu }=f^{\prime }(R_{0})\tau _{LL}^{\mu \nu
}+\frac{1}{6\kappa } \{f^{\prime
}(R_{0})R_{0}-f(R_{0})\}\frac{\partial }{\partial x^{\lambda }}
(g^{\mu \nu }x^{\lambda }-g^{\mu \lambda }x^{\nu }),  \label{9}
\end{equation}%
where $\tau _{LL}^{\mu \nu }$ is the Landau-Lifshitz EMC in GR and
$\kappa =8\pi G$. We can calculate EMD in the framework of $f(R)$
theory of any metric tensor which has constant scalar curvature.
Its 00-component represents the energy density and is given by the
following equation
\begin{equation}
\tau ^{00}=f^{\prime }(R_{0})\tau _{LL}^{00}+\frac{1}{6\kappa }(f^{\prime
}(R_{0})R_{0}-f(R_{0}))(\frac{\partial }{\partial x^{i}}g^{00}x^{i}+3g^{00}),
\label{10}
\end{equation}%
where $\tau _{LL}^{00}$ represents the sum of energy-momentum
tensor and the energy-momentum pseudo tensor and is given by
\begin{equation}
\tau _{LL}^{00}=(-g)(T^{00}+t_{LL}^{00})  \label{11}
\end{equation}%
and
\begin{equation}
T^{00}=\frac{1}{\kappa }(R^{00}-\frac{1}{2}g^{00}R),  \label{12}
\end{equation}%
where $R$ is the Ricci scalar and $t_{LL}^{00}$ can be evaluated
from the following expression
\begin{eqnarray}
t_{LL}^{00} &=&\frac{1}{2\kappa }[(2\Gamma _{\alpha \beta
}^{\gamma }\Gamma _{\gamma \delta }^{\delta }-\Gamma _{\alpha
\delta }^{\gamma }\Gamma _{\beta \gamma }^{\delta }-\Gamma
_{\alpha \gamma }^{\gamma }\Gamma _{\beta \delta }^{\delta
})(g^{\mu \alpha}g^{\nu\beta}-g^{\mu\nu}g^{\alpha\beta}) \nonumber\\
&+&g^{\mu \alpha }g^{\beta \gamma }(\Gamma _{\alpha \delta }^{\nu }\Gamma
_{\beta \gamma }^{\delta }+\Gamma _{\beta \gamma }^{\nu }\Gamma _{\alpha
\delta }^{\delta }-\Gamma _{\gamma \delta }^{\nu }\Gamma _{\alpha \beta
}^{\delta }-\Gamma _{\alpha \beta }^{\nu }\Gamma _{\gamma \delta }^{\delta })\nonumber\\
&+&g^{\nu \alpha }g^{\beta \gamma }(\Gamma _{\alpha \delta }^{\mu }\Gamma
_{\beta \gamma }^{\delta }+\Gamma _{\beta \gamma }^{\mu }\Gamma _{\alpha
\delta }^{\delta }-\Gamma _{\gamma \delta }^{\mu }\Gamma _{\alpha \beta
}^{\delta }-\Gamma _{\alpha \beta }^{\mu }\Gamma _{\gamma \delta }^{\delta })\nonumber\\
&+&g^{\alpha \beta }g^{\gamma \delta }(\Gamma _{\alpha \gamma }^{\mu }\Gamma
_{\beta \delta }^{\nu }-\Gamma _{\alpha \beta }^{\mu }\Gamma _{\gamma \delta
}^{\nu })].  \label{13}
\end{eqnarray}%

\section{Spatially Homogeneous Rotating Spacetimes Solution}

In this section, we solve the vacuum field equations of $f(R)$
theory for the metric representing  the spatially homogeneous
rotating
spacetimes by using metric approach and the condition of constant scalar curvature, i.e., $%
(R=constant)$. The line element representing the spatially
homogeneous rotating spacetimes is given by
\begin{equation}
ds^{2}=dt^{2}-dr^{2}-A(r)d\phi ^{2}-dz^{2}+2B(r)dtd\phi ,  \label{14}
\end{equation}%
where $A(r)$ and $B(r)$ are arbitrary functions of $r$. This
metric represents five spacetimes \cite{[30]}-\cite{[31]}, which
can be achieved by choosing particular values of the metric
functions $A$ and $B$.\newline The corresponding Ricci scalar is
given by
\begin{equation}
R=\frac{4ABB^{^{\prime \prime }}+4B^{3}B^{\prime \prime
}-4A^{\prime }B^{\prime }B +3AB^{\prime 2}-B^{2}B^{\prime
2}+2AA^{\prime \prime } +2A^{\prime \prime }B^{2}-A^{\prime
2}}{2(A+B^{2})^{2}}, \label{15}
\end{equation}
where prime denotes the derivative with respect to $r$.
Eq.(\ref{5}) implies that
\begin{equation}
f(R)=\frac{3 \Box F(R)+F(R)R}{2}. \label{16}
\end{equation}
Making use of this value of $f(R)$ in Eq.(\ref{2}) along with the
substitution $T_{\mu\nu}=0$ (for vacuum case), we reached at
\begin{equation}
\frac{F(R)R_{\mu \nu }-\nabla _{\mu }\nabla _{\nu }F(R)}{g_{\mu
\nu }}=\frac{F(R)R-\Box F(R)}{4}.  \label{17}
\end{equation}
Since the metric Eq.(\ref{14}) depends only upon r so
Eq.(\ref{17}) can be viewed as the set of differential equations
for $F(R)$, $A(r)$ and $B(r)$. Eq.(\ref{17}) can be reduced to the
following combination
\begin{equation}
A_{\mu }=\frac{F(R)R_{\mu \mu }-\nabla _{\mu }\nabla _{\mu }F(R)}{g_{\mu \mu
}}.  \label{18}
\end{equation}%
This combination is independent of of the index $\mu $ and hence
$A_{\mu }-A_{\nu }=0$ for all $\mu $ and $\nu $. Thus,
$A_{0}-A_{1}=0$ results that
\begin{eqnarray}
&&(4B^{2}B^{\prime 2}-4ABB^{\prime \prime }-4B^{\prime \prime
}B^{3}+4A^{\prime }B^{\prime }B-2AA^{\prime \prime }
-2B^{2}A^{\prime \prime }+A^{\prime
2})F\nonumber\\
&&-4(A+B^{2})^{2}F^{\prime \prime }=0.  \label{19}
\end{eqnarray}%
Similarly, for all other possible combinations, i.e., $A_{0}-A_{2}=0$, $%
A_{0}-A_{3}=0$, $A_{1}-A_{2}=0$, $A_{1}-A_{3}=0$ and
$A_{2}-A_{3}=0$, we have the following independent equations
\begin{eqnarray}
(2AA^{\prime \prime }+2B^{2}A^{\prime \prime }+2A^{\prime
}B^{\prime }B-A^{\prime 2})F-2A^{\prime }(A+B^{2})F^{\prime
}&=&0,
\\
\label{20} (\frac{B^{\prime 2}}{2(A+B^{2})})F&=&0,
\label{21}
\\
(4A^{2}BB^{\prime \prime }+4B^{3}B^{\prime \prime }-2A^{\prime
}B^{\prime }AB-2AA^{\prime \prime }B^{2}-2A^{\prime \prime
}B^{4}\nonumber\\+2A^{\prime }B^{\prime }B^{3}+A^{\prime }B^{2})F
 -2A^{\prime }(A+B^{2})^{2}F^{\prime }+4A(A+B^{2})^{2}F^{\prime
\prime }&=&0, \label{22}
\\
(2AB^{\prime 2}+4ABB^{\prime \prime }+4B^{3}B^{\prime \prime
}-4A^{\prime }B^{\prime }B-2B^{\prime 2}B^{2}+2AA^{\prime \prime
}\nonumber\\+2A^{\prime \prime }B^{2}-A^{\prime 2})F
-4(A+B^{2})^{2}F^{\prime \prime }&=&0,  \label{23}
\\
(2AB^{\prime 2}+2AA^{\prime \prime }-A^{\prime 2}-2A^{\prime
}B^{\prime }B+2A^{\prime \prime
}B^{2})F\nonumber\\
-2\frac{A^{\prime }}{A}(A+B^{2})F^{\prime }&=&0. \label{24}
\end{eqnarray}%
These are the six non-linear ordinary differential equations
involving three unknown variables $A$,$B$ and $F$. We use the
condition of constant curvature to solve these equations as given
in the next subsection.

\subsection{Constant Curvature Solution}

For constant curvature solution, that is, $R=R_{0}$, it is obvious
that the first and second derivatives of $F(R)=\frac{df(R)}{dR}$
will always vanish, that is,
\begin{equation}
F^{\prime }(R_{0})=0=F^{\prime \prime }(R_{0}).  \label{25}
\end{equation}
In view of the Eq.(\ref{25}), the Eqs. (\ref{19})-(\ref{24})
reduce to
\begin{eqnarray} \label{26}
4B^{2}B^{\prime 2}-4ABB^{\prime \prime }-4B^{\prime \prime
}B^{3}+4A^{\prime }B^{\prime }B-2AA^{\prime\prime}
\nonumber\\
-2B^{2}A^{\prime \prime }+A^{\prime 2}&=&0,
\\
2AA^{\prime \prime }+2B^{2}A^{\prime \prime }+2A^{\prime
}B^{\prime }B-A^{\prime 2}&=&0,  \label{27}
\\
B^{\prime 2}&=&0, \label{28}
\\
4A^{2}BB^{\prime \prime }+4B^{3}B^{\prime \prime }-2A^{\prime
}B^{\prime }AB-2AA^{\prime \prime }B^{2}-2A^{\prime \prime
}B^{4}\nonumber\\+2A^{\prime }B^{\prime }B^{3}+A^{\prime
}B^{2}&=&0, \label{29}
\\
2AB^{\prime 2}+4ABB^{\prime \prime }+4B^{3}B^{\prime \prime
}-4A^{\prime }B^{\prime }B-2B^{\prime 2}B^{2}+2AA^{\prime \prime
}\nonumber\\+2A^{\prime \prime }B^{2}-A^{\prime 2}&=&0,
\label{30}
\\
2AB^{\prime 2}+2AA^{\prime \prime }-A^{\prime 2}-2A^{\prime
}B^{\prime }B+2A^{\prime \prime }B^{2}&=&0.  \label{31}
\end{eqnarray}%
Eq.(\ref{28}) implies that
\begin{equation}
B=constant=c_{1}~~ (say). \label{32}
\end{equation}%
After inserting this value of $B$, the Eqs.(\ref{26})-(\ref{27})
and (\ref{29})-(\ref{31}) reduce to a single equation
\begin{equation}
2AA^{\prime \prime }+2c_{1}^{2}A^{\prime \prime }-A^{\prime 2}=0.
\label{33}
\end{equation}%
This is a second order ordinary differential equation and its
non-trivial solution is given (by using the software Maple) as
\begin{equation}
A(r)=\frac{1}{4}c_{2}^{2}r^{2}+\frac{1}{2}c_{2}c_{3}r+\frac{1}{4}%
c_{3}^{2}-c_{1}^{2},  \label{34}
\end{equation}%
where $c_{2}$ and $c_{3}$ are constants. It is noticed that
$A=constant$  is a trivial solution of the Eq.(\ref{33}). Now,
renaming the constants, the Eq.(\ref{34}) takes the form
\begin{equation}
A(r)=c_{4}r^{2}+c_{5}r+c_{6},  \label{35}
\end{equation}%
where $c_{4}=\frac{1}{4}c_{2}^{2}$, $c_{5}=\frac{1}{2}c_{2}c_{3}$ and $%
c_{6}=\frac{1}{4}c_{3}^{2}-c_{1}^{2}$. Hence, the corresponding
solution turns out to be
\begin{equation}
ds^{2}=dt^{2}-dr^{2}-(c_{4}r^{2}+c_{5}r+c_{6})d\phi
^{2}-dz^{2}+2c_{1}dtd\phi \label{36}
\end{equation}
It is mentioned here that the Ricci scalar evaluated for this
solution vanishes identically, i.e.,
\begin{equation}
R=0  \label{37}
\end{equation}

\section{Energy Density of the Spatially Homogeneous Rotating Solution}

In this section, we compute energy density of the spatially
homogeneous rotating solution (\ref{36}), which is  obtained in in
the framework of $f(R)$ theory of gravity in the last section. For
this purpose, we use generalized Landau-Lifshitz EMC in the
context of $f(R)$ gravity. Substituting the value of $g^{00}$, the
Eq.(\ref{10}), takes the form
\begin{equation}
\tau ^{00}=f^{\prime }(R_{0})\tau _{LL}^{00}+\frac{1}{2\kappa
}(f^{\prime }(R_{0})R_{0}-f(R_{0})).  \label{39}
\end{equation}
Now, by evaluating $T^{00}$ and $t^{00}_{LL}$ from Eqs.(\ref{12})
and (\ref{13}) respectively and then using in Eq.(\ref{11}), $\tau
_{LL}^{00}$ turns out to be
\begin{equation}
\tau _{LL}^{00}=\frac{1}{4\kappa }c_{2}^2. \label{40}
\end{equation}
Using this value of $\tau _{LL}^{00}$ in Eq.(\ref{39}), the
00-component of the Landau-Lifshitz EMC takes the form
\begin{equation}
\tau ^{00}=f^{\prime }(R_{0})\frac{1}{4\kappa }(c_{2}^2)+%
\frac{1}{2\kappa }(f^{\prime }(R_{0})R_{0}-f(R_{0})). \label{41}
\end{equation}
Finally, we will use suitable $f(R)$ model to calculate this
component completely. It is revealed here that there must be some
restrictions to choose the $f(R)$ model when $R=0$ because if the
model involves the logarithmic function of Ricci scalar $R$ or a
linear superposition of \ $R^{-n}$, where n is positive integer,
then we can not evaluate this EMC. Thus, we consider the following
model
\begin{equation}
f(R)=R+\epsilon R^{2},  \label{42}
\end{equation}
where $\epsilon $ is a positive real number. Accordingly, the
00-component of generalized Landau-Lifshitz EMC results as
\begin{equation}
\tau ^{00}=\frac{1}{4\kappa }c_{2}^2. \label{43}
\end{equation}
Furthermore, the stability condition for this $f(R)$
model$\ $%
\begin{equation}
\frac{1}{\epsilon (1+2\epsilon R_{0})}=\frac{1}{\epsilon }>0
\label{44}
\end{equation}
holds. The condition of constant scalar curvature is also
satisfied for this $f(R)$ model.

\section{Summary and Conclusion}

This paper has mainly two parts: In first part, we have explored
the spatially homogeneous rotating solution in $f(R)$ theory of
gravity. For this purpose, we solve the field equations of $f(R)$
gravity for the metric representing the spatially homogeneous
rotating spacetimes. It is found that there arise two solutions,
one trivial (when the metric functions $A(r)$ and $B(r)$ become
constant) and other one is the non-trivial solution. It is
interesting to mention here that the Ricci scalar $R=0$ vanishes
in both the cases.

The second part of the paper contains the energy density component
of the non-trivial solution. For this purpose, we used the
generalized Landau-Lifshitz EMC in the framework of $f(R)$ theory
of gravity for a suitable $f(R)$ model. It has been shown that the
particularly selected $f(R)$ model satisfies the condition of
constant scalar curvature, which is the mandatory constraint for
the validity of $f(R)$ models. Further, we have discussed the
stability condition for this model. It is mentioned here that the
energy density for this model turns out to be constant as given in
Eq.(\ref{43}). It is anticipated that such solutions may provide
an entrance towards the solution of dark energy and dark matter
problems.

\bigskip

\bigskip\ \ \ \ \ \ \ \ \ \ \ \ \ \

\end{document}